\documentclass[%
 aip,
 jmp,%
 amsmath,amssymb,
 reprint,%
]{revtex4-2}

\usepackage{graphicx}
\usepackage{dcolumn}
\usepackage{bm}
\usepackage{float}
\usepackage[caption = false]{subfig}

\usepackage[utf8]{inputenc}
\usepackage[T1]{fontenc}
\usepackage{mathptmx}
\usepackage{etoolbox}

\begin{document}

\preprint{AIP/123-QED}

\title[Role of structural biaxiality on the phase behaviour of chiral liquid crystals]{Role of structural biaxiality on the phase behaviour of chiral liquid crystals}

\author{Sayantan Mondal}
 \email{sayantan1sept1995@gmail.com}
\author{Jayashree Saha}
 \email{jsphy@caluniv.ac.in}
\affiliation{ 
Department of Physics, University of Calcutta, 92, APC Road, Kolkata 700009, India
}%


\date{\today}

\begin{abstract}
We report a computer simulation study on the effect of molecular structural biaxiality in the phase formation of chiral molecules. In this study, we have done coarse-grained modeling to observe self-assembled phase behavior. In our molecular dynamics simulation study we varied both the chiral interaction strength and molecular biaxiality. Uniaxial molecules give rise to cholesteric phase, blue phase whereas molecular biaxiality favours cholesteric phase. At higher chirality, small chiral domains are formed creating twisted cylindrical networks with each cylinder having elliptical cross-sections instead of circular nature as found in uniaxial systems. The value of cholesteric pitch decreases when chirality and molecular biaxiality becomes higher. Coaction of biaxiality and chirality is crucial for fabricating liquid crystal materials with optical properties suitable for displays, sensors and chiral photonic devices.
\end{abstract}

\maketitle

%

\section{\label{sec:level1}Introduction}

Chirality is an absence of mirror symmetry, where due to broken improper rotational symmetry chiral mirror images can not be superimposed giving rise to a handedness. 
Chiral molecules can lead to helical superstructures in the liquid crystalline systems. These mesophases which are fluidic in nature, also shows a local orientational order. Experimental studies{\cite{Goodby,Grelet}} revealed different kinds of helical phases like cholesteric phase, smectic $C^{*}$, smectic $A^{*}$ and also some frustrated mesophase like blue phase, twist grain boundary phases etc. Goossens{\cite{Goossens}} first proposed a molecular theory of the cholesteric phase taking into account the induced dipole-dipole interaction alongwith the dipole-quadrupole interaction. Molecular-statistical theory{\cite{Schroder1,Schroder2,meer}} of Van der Meer et al{\cite{van der meer}} assumed the chiral molecules to behave as if they were cylindrically symmetric. Computer simulation studies are crucial to understand the connection between the microscopic details of molecular level interaction and the macroscopic helicity of the liquid crystalline mesophases. The difference between the directions along and perpendicular to the helical director axis of a cholesteric phase was first noticed by Priest and Lubensky{\cite{Priest}}. Chirality induced biaxial effects were studied{\cite{Yaniv,Kroin,Lin-Liu,Harris}} and a model developed by Dhakal et al{\cite{Dhakal}} showed how cholesteric twist and biaxial order could have effects{\cite{Reddy,Gorecka,Archer,Ogawa}} on one another. Statistical mechanical studies on the effect of biaxial orderings on chiral pitch has been done{\cite{hosino1,hosino2}}. The phase transition between cholesteric-$I$ to cholesteric-$II$ has been studied{\cite{Brand}}. Since the discovery of a biaxial nematic phase in thermotropic liquid crystals{\cite{Acharya,Madsen}}, the application of the biaxial liquid crystals for fast-switching display devices{\cite{Kumar}} has grown. The applications of Cholesteric liquid crystals{\cite{Coates,Gollapelli,Senyuk,Moreira,Schmidtke,Cao,Smalyukh,Wu}} include displays, switchable gratings, sensors, lasers, optical anti-counterfeiting etc. 
\newline A lattice model simulation study for a system of chiral molecules giving full rotational degrees of freedom was reported by Saha et al{\cite{saha}}. Coarse grained Monte carlo simulation of uniaxial chiral calamitic molecules with canonical ensemble(NVT) was studied by Memmer et al{\cite{memmer,memmer_1}}, where they considered Gay-Berne{\cite{gay}} potential with an additive chiral interaction potential energy. Screw-like nematic phase has recently been observed in a simulation study of helical particles{\cite{Cinacchi}}. Phase transitional behavior of uniaxial chiral molecules with embedded terminal point dipoles was studied by Paul et al{\cite{paul2}}, which revealed that the gradual change of the strength of the chiral interaction gave cholesteric and blue phases and also some other novel chiral phases{\cite{paul3}}. 
\newline To account for the true ellipsoidal geometry of molecules having shape of any kind of spheroids, the elliptic contact function(ECF) was used{\cite{Perram,Zheng2,Patey,Paramonov}}. Ellipsoid contact potential(ECP) developed by Perram et al{\cite{Perram_et_al}} calculated the contact points of the molecular ellipsoids. But this ECP did not take into account the configuration dependent energy anisotropies. It was introduced later by Saha{\cite{saha2}} with a soft ellipsoid contact potential for uniaxial ellipsoidal molecules and in their later work{\cite{saha3}} they generalized the model potential by including biaxial contribution of the distance and energy of each molecular ellipsoids. 
\\ In this work, we are presenting a molecular dynamics simulation study of a system of biaxial molecular ellipsoids where each molecule has a chirality. Our system consists of a single type of prolate ellipsoids with biaxiality which means they have two axes of rotational symmetry or they have orientational order along mutually perpendicular axes. 

\section{Model}

In this work, we have used a coarse-grained single-site NVT molecular dynamics(MD) simulation using biaxial ellipsoid contact potential{\cite{saha3}} for prolate ellipsoidal molecules which have chirality{\cite{paul2,paul3}}. 
\\The total interaction pair potential acting between two neighbouring $i$ th and $j$ th molecule is given by,
\begin{widetext}
\begin{eqnarray}
U =\hspace{+0.2cm} &&4\epsilon(\hat{a}_{i},\hat{a}_{j},\hat{r}_{ij})\left[\left(\frac{\sigma_0}{r_{ij}-\sigma(\hat{a}_{i},\hat{a}_{j},\hat{r}_{ij})+\sigma_0}\right)^{12}-\left(\frac{\sigma_0}{r_{ij}-\sigma(\hat{a}_{i},\hat{a}_{j},\hat{r}_{ij})+\sigma_0}\right)^{6}\right] \nonumber \\ -\hspace{+0.2cm} &&c4\epsilon(\hat{a}_{i},\hat{a}_{j},\hat{r}_{ij})\left(\frac{\sigma_0}{r_{ij}-\sigma(\hat{a}_{i},\hat{a}_{j},\hat{r}_{ij})+\sigma_0}\right)^{7}[(\hat{a}_{i3}\times\hat{a}_{j3})\cdot{\hat{r}_{ij}}](\hat{a}_{i3}\cdot\hat{a}_{j3}) 
\end{eqnarray}
\end{widetext}
Here, the first term considers biaxial ellipsoid contact potential, which generates van-der-Waals force which is dispersive in nature. 
$c$ is the chiral strength parameter. The second term is a pseudoscalar term consisting of a box product, which produces an intermolecular torque of a particular handedness depending upon the sign of chiral strength parameter. The chiral strength parameter $c$ is zero for achiral molecules and it is non-zero for chiral molecules.
\\ In the above equation $\hat{a}_{i3}$ and $\hat{a}_{j3}$ are the unit vectors along semi-major axis(molecular z-axis) of molecules $i$ and $j$ respectively and ${r}_{ij}$ is the inter-molecular distance between the center of mass of the molecular ellipsoids. Also, $\hat{a}_{i}$ and $\hat{a}_{j}$ are orthogonal rotation matrices for transformation of molecules $i$ and $j$ from laboratory frame to body frame coordinates and molecular orientational information is stored into them. Here, $\sigma_0$ is the length of smallest semi-axis of each ellipsoidal molecule. The distance of closest approach between the molecules is $\sigma(\hat{a}_{i},\hat{a}_{j},\hat{r}_{ij})$, which depends on the orientations and intermolecular separation between them for our anisotropic system.
\\The form of the anisotropic range parameter $\sigma$ is given by,
\begin{eqnarray}
\sigma^{-2} = \lambda_{D}(1-\lambda_{D})(\hat{r}^T.{\bf{H_{D}}}^{-1}.\hat{r}) \label{eq:2}
\end{eqnarray}
Where, ${\mathbf{H_{D}}}$ is an affine combination of two matrices. 
\begin{eqnarray} 
{\mathbf{H_{D}}} = \lambda_{D}{\bf{M}} + (1-\lambda_{D}){\bf{N}}
\end{eqnarray}
The matrices are defined as,
\begin{equation}
{\bf{M}} = \hat{a}_{i}^{T}{\bf{S}}^2\hat{a}_{i}\hspace{0.2cm};\hspace{0.2cm}{\bf{N}} = \hat{a}_{j}^{T}{\bf{S}}^2\hat{a}_{j}
\end{equation}
The diagonal shape matrix ${\bf{S}}$ has the form,
\begin{equation}
{\bf{S}} = \begin{pmatrix}
\sigma_x & 0 & 0 \\
0 & \sigma_y & 0 \\
0 & 0 & \sigma_z
\end{pmatrix}
\end{equation}
Where, $\sigma_x$, $\sigma_y$ and $\sigma_z$ are the values of range parameters when the neighbouring molecules are at side-side(s-s), width-width(w-w) and end-end(e-e) configurations respectively. For our study we have considered systems for which $\sigma_x<\sigma_y<\sigma_z$ and $\sigma_x=\sigma_0$. The scalar $\lambda_{D}$ is a shape-related adjustable parameter, which is evaluated by optimizing equation \eqref{eq:2} using Brent's method{\cite{Brent}}, it takes value between $\in[0,1]$.
\\The total anisotropic energy function of our system takes the form,
\begin{eqnarray}
\epsilon(\hat{a}_{i},\hat{a}_{j},\hat{r}_{ij}) = \epsilon_0 \epsilon^{\nu}_1(\hat{a}_{i},\hat{a}_{j}) \epsilon^{\mu}_2(\hat{a}_{i},\hat{a}_{j},\hat{r}_{ij})
\end{eqnarray}
Where, $\epsilon_0$ is a constant term and $\mu$, $\nu$ are adjustable parameters for well-depth variations in various different compounds. $\epsilon_1(\hat{a}_{i},\hat{a}_{j})$ depends only on orientation{\cite{zanoni}}, but $\epsilon_2(\hat{a}_{i},\hat{a}_{j},\hat{r}_{ij})$ depends on orientations as well as on the intermolecular separation between the molecules. 
\\Defining a diagonal interaction energy matrix, 
\begin{equation}
{\bf{\gamma}} = \begin{pmatrix}
(\frac{\epsilon_0}{\epsilon_x})^{\frac{1}{\mu}} & 0 & 0 \\
0 & (\frac{\epsilon_0}{\epsilon_y})^{\frac{1}{\mu}} & 0 \\
0 & 0 & (\frac{\epsilon_0}{\epsilon_z})^{\frac{1}{\mu}}
\end{pmatrix}
\end{equation}
Here, $\epsilon_x,\epsilon_y,\epsilon_z$ are the values of the well depth for s-s, w-w and e-e configurations respectively. For our systems we have considered $\epsilon_x>\epsilon_y>\epsilon_z$ and $\epsilon_x=\epsilon_0$ for our model.
The position dependent part of the energy function takes the form,
\begin{equation}
\epsilon_{2}(\hat{a}_{i},\hat{a}_{j},\hat{r}) = \lambda_{E}(1-\lambda_{E})(\hat{r}^T.{\bf{H_E}}^{-1}.\hat{r}) \label{eq:8}
\end{equation}
Here, the expression of ${\bf{H_E}}$ is given by, 
\begin{equation}
{\bf{H_E}} = \lambda_E{\bf{A}}+(1-\lambda_E){\bf{B}} \label{eq:9}
\end{equation}
where, $\lambda_E$ is an energy-related adjustable parameter such that $\lambda_E\in[0,1]$, which is evaluated by optimizing equation \eqref{eq:8} using Brent's method{\cite{Brent}}.
The matrices in equation \eqref{eq:9} has the form,
\begin{equation}
{\bf{A}} = \hat{a}_{i}^{T}{\bf{\gamma}}\hat{a}_{i}\hspace{0.2cm};\hspace{0.2cm}{\bf{B}} = \hat{a}_{j}^{T}{\bf{\gamma}}\hat{a}_{j}
\end{equation}

And $\epsilon_1$ is given by{\cite{zanoni}},
\begin{equation}
\epsilon_{1}(\hat{a}_{i},\hat{a}_{j}) = \frac{1}{8}\sqrt{\frac{2\sigma_{y}}{\sigma_{x}}}\sigma^3_0\left[(\frac{\sigma_{y}}{\sigma_{x}})+(\frac{\sigma_{z}}{\sigma_{x}})^{2}\right]{\left|{\bf{H^{\prime}_E}}\right|^{-\frac{1}{2}}}
\end{equation}
Where, ${\bf{H^{\prime}_E}}\hspace{-0.1cm}=\hspace{-0.1cm}({\bf{A}}+{\bf{B}})$. In the limiting case, when $\lambda_D=\lambda_E=\frac{1}{2}$ this model reduces down into the standard Gaussian Overlap Potential(GOP) models\cite{berne,gay}. However for the true ellipsoidal geometry, $\lambda$'s values are determined numerically at every time-step by the optimization of corresponding object functions described in equations \eqref{eq:2} and \eqref{eq:8}. Here, the box product term is denoted by ${\bf{b_p}}=\{(\hat{a}_{i3}\times\hat{a}_{j3})\cdot{\hat{r}_{ij}}\}(\hat{a}_{i3}\cdot\hat{a}_{j3})$. Defining vectors ${\bf{\kappa}}$ \& ${\bf{\kappa_{E}}}$ by the relations ${\mathbf{H_{D}}}\cdot{\bf{\kappa}}={\mathbf{r}}$ \& ${\mathbf{H_{E}}}\cdot{\bf{\kappa_{E}}}={\mathbf{r}}$. The expressions{\cite{Germano}} of forces and torques are given by,
\begin{widetext}
\begin{eqnarray}
\vec{f} &=& \frac{8{\epsilon_0}{\epsilon^{\nu}_1}{\epsilon^{\mu}_2}}{\sigma_0r^{2}}[3r^{2}(2\rho^{-13}-\rho^{-7})\hat{r} + 3{\sigma^{3}}{\lambda_D(1-\lambda_D)}(2\rho^{-13}-\rho^{-7})\{{\bf{\kappa}}-(\hat{r}.{\bf{\kappa}})\hat{r}\} \nonumber \\ &-& {\lambda_E(1-\lambda_E)}{\mu}{\epsilon^{-1}_2}{\sigma_0}(\rho^{-12}-\rho^{-6})\{{\bf{\kappa_{E}}}-(\hat{r}.{\bf{\kappa_{E}}})\hat{r}\}] \nonumber \\ &+& \frac{4{c}{\epsilon_0}{\epsilon^{\nu}_1}{\epsilon^{\mu}_2}}{\sigma_0r^{2}}[-7r^{2}{\rho^{-8}}{\bf{b_p}}\hat{r} - 7{\sigma^{3}}{\lambda_D(1-\lambda_D)}{\rho^{-8}}{\bf{b_p}}\{{\bf{\kappa}}-(\hat{r}.{\bf{\kappa}})\hat{r}\} \nonumber \\ &+& 2{\lambda_E(1-\lambda_E)}{\mu}{\epsilon^{-1}_2}{\sigma_0}{\rho^{-7}}{\bf{b_p}}\{{\bf{\kappa_{E}}}-(\hat{r}.{\bf{\kappa_{E}}})\hat{r}\} \nonumber \\ &+& {\rho^{-7}}{\sigma_0r^{2}}\{\frac{(\hat{a}_{i3}\times\hat{a}_{j3})\cdot(\hat{a}_{i3}\cdot\hat{a}_{j3})}{ \lvert \vec{r} \rvert }-\frac{\vec{r}}{r^{2}}{\bf{b_p}}\}] 
\end{eqnarray}
\begin{eqnarray}
\vec{{\tau}}_{A} &=& \frac{8{\epsilon_0}{\epsilon^{\nu}_1}{\epsilon^{\mu}_2}}{r^{2}}[-3{\sigma^{3}}{\lambda^{2}_D(1-\lambda_D)}\frac{(2\rho^{-13}-\rho^{-7})}{\sigma_0}\{{\bf{\kappa}}\cdot{\bf{M}}\times{\bf{\kappa}}\} \nonumber \\ &-& {\lambda^{2}_E(1-\lambda_E)}{\mu}{\epsilon^{-1}_2}(\rho^{-12}-\rho^{-6})\{{\bf{\kappa_{E}}}\cdot{\bf{A}}\times{\bf{\kappa_{E}}}\} + \frac{1}{2}\nu{r^{2}}(\rho^{-12}-\rho^{-6}){\times} \{\hat{a}_{i}^{T}\cdot({\bf{A}}+{\bf{B}})^{-1}\cdot\hat{a}_{i}\}] \nonumber \\ &+& \frac{4{c}{\epsilon_0}{\epsilon^{\nu}_1}{\epsilon^{\mu}_2}}{\sigma_0r^{2}}[7{\sigma^{3}}{\lambda^{2}_D(1-\lambda_D)}{\rho^{-8}}{\bf{b_p}}\{{\bf{\kappa}}\cdot{\bf{M}}\times{\bf{\kappa}}\} - 2{\lambda^{2}_E(1-\lambda_E)}{\mu}{\epsilon^{-1}_2}{\sigma_0}{\rho^{-7}}{\bf{b_p}}\{{\bf{\kappa_{E}}}\cdot{\bf{A}}\times{\bf{\kappa_{E}}}\} \nonumber \\ &-& {\sigma_0r^{2}}{\nu}{\rho^{-7}}{\bf{b_p}}\{\hat{a}_{i}^{T}\cdot({\bf{A}}+{\bf{B}})^{-1}\cdot\hat{a}_{i}\} \nonumber \\ &+& {\rho^{-7}}{\sigma_0r^{2}}(\hat{a}_{i3}\times\{\frac{({\hat{a}_{j3}}\times{\vec{r}})(\hat{a}_{i3}\cdot\hat{a}_{j3})}{\vec{r}} + \{(\hat{a}_{i3}\times\hat{a}_{j3})\cdot{\hat{r}_{ij}}\}{\hat{a}_{j3}}\})] 
\end{eqnarray}
\end{widetext}
Where, $\vec{{\tau}}_{A}$ is the torque on one molecule due to the other.

\section{Simulation Details and Results}

\begin{figure}
\subfloat[set-1 biaxiality]{\includegraphics[angle=-90,width=4.25cm]{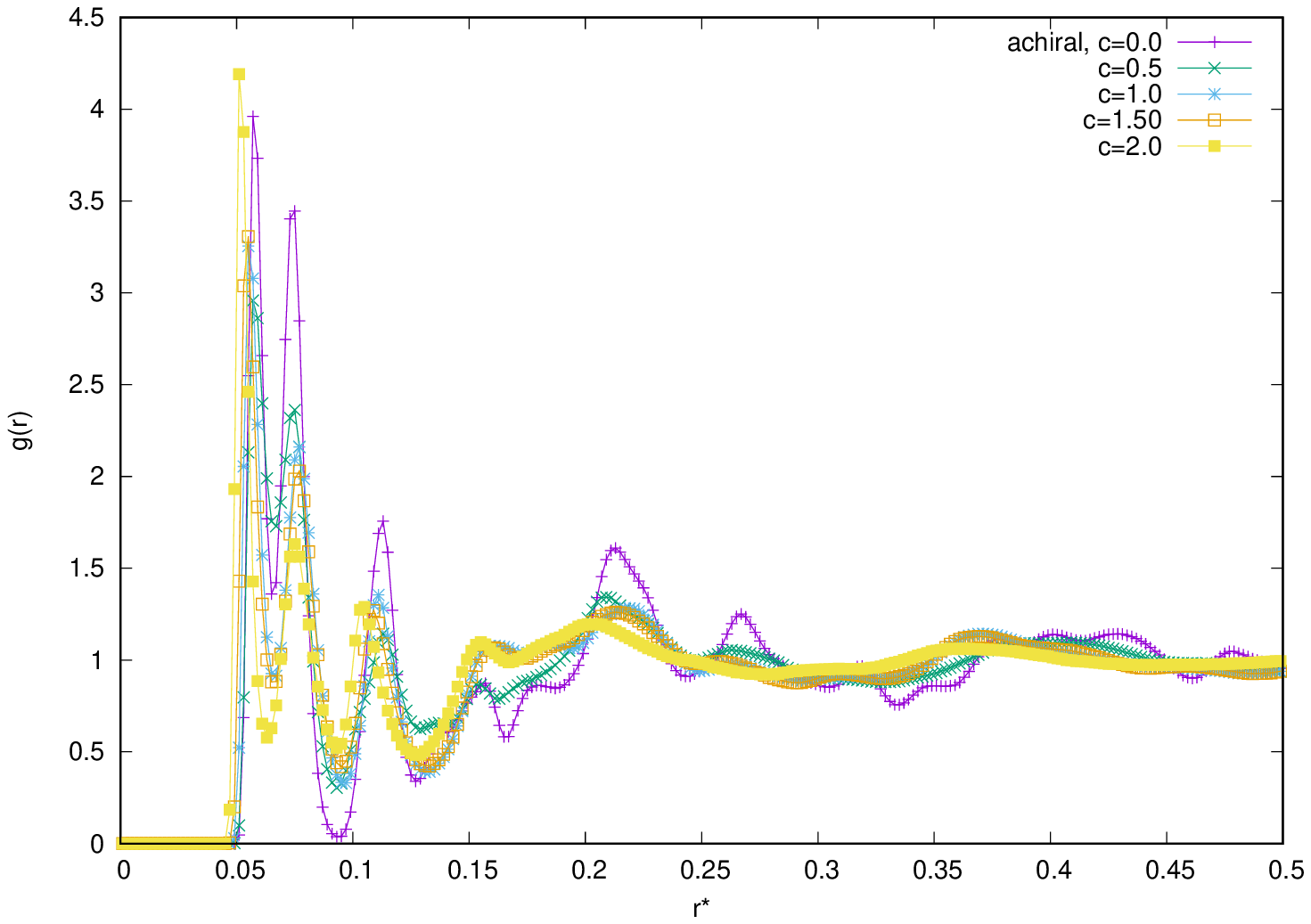}}
\subfloat[set-2 biaxiality]{\includegraphics[angle=-90,width=4.25cm]{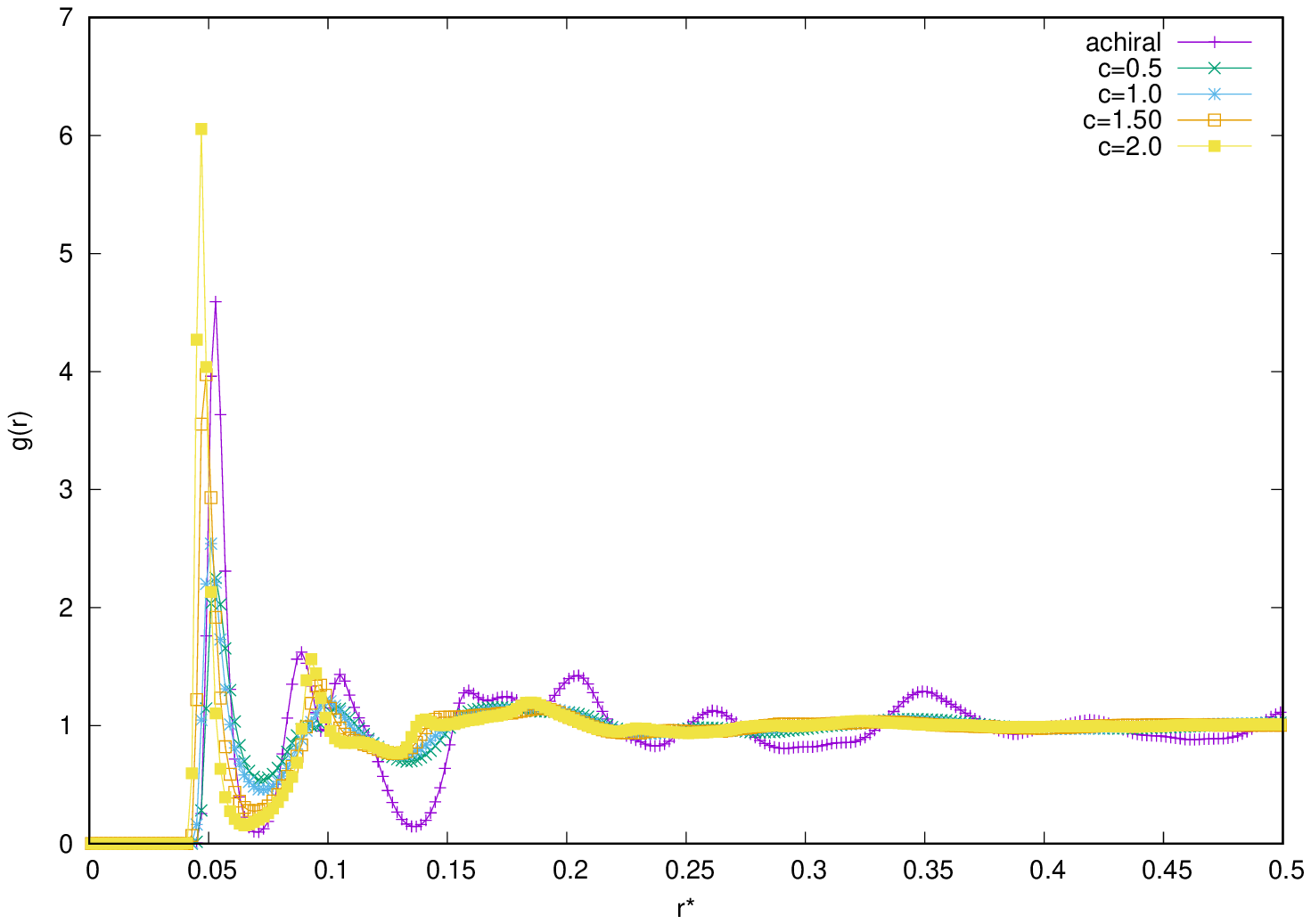}}
\caption{\label{fig:3} Plot of g(r) vs r* at different chiral strength}
\end{figure}

\begin{figure}
\subfloat[set-1 biaxiality]{\includegraphics[angle=-90,width=4.25cm]{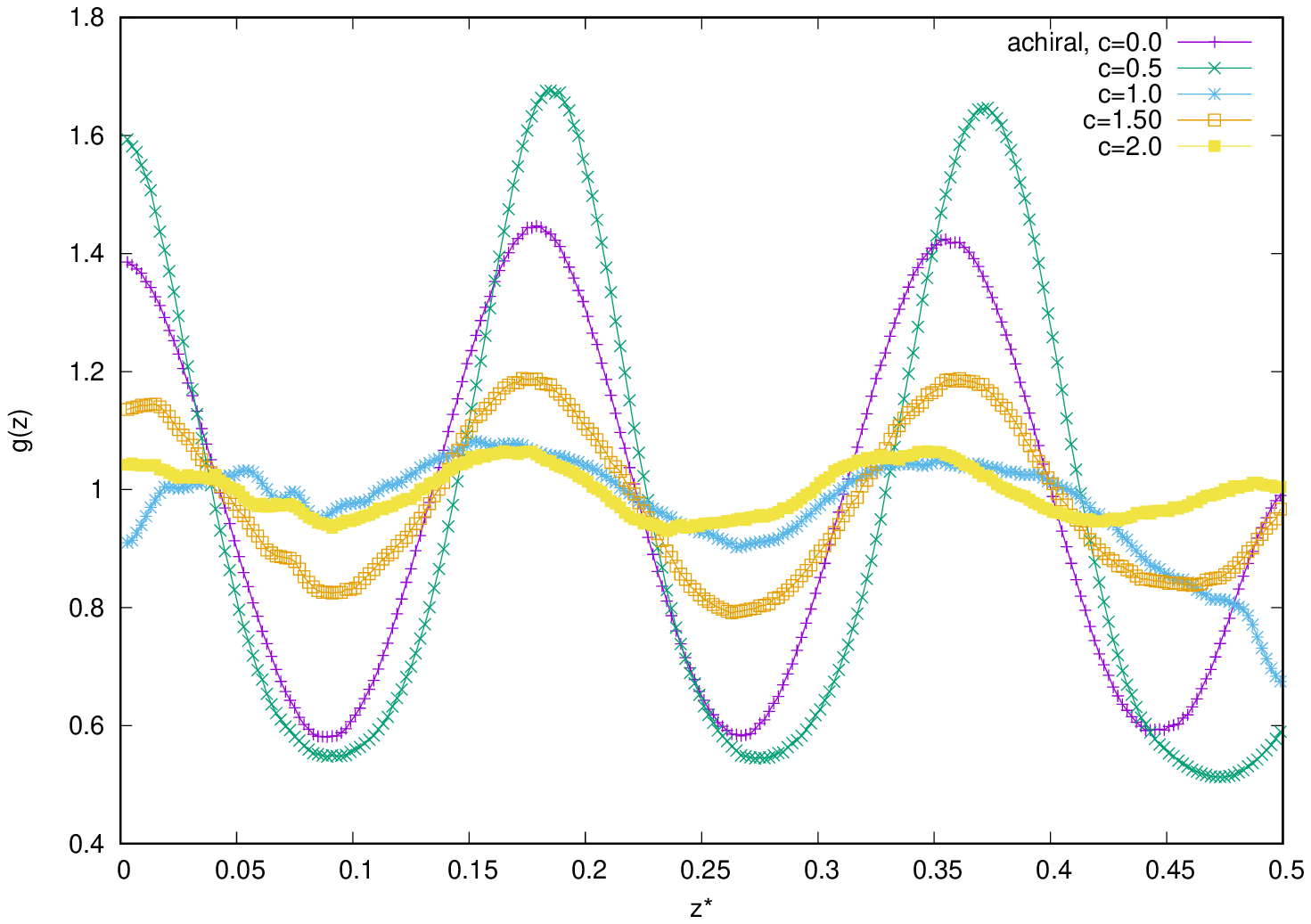}}
\subfloat[set-2 biaxiality]{\includegraphics[angle=-90,width=4.25cm]{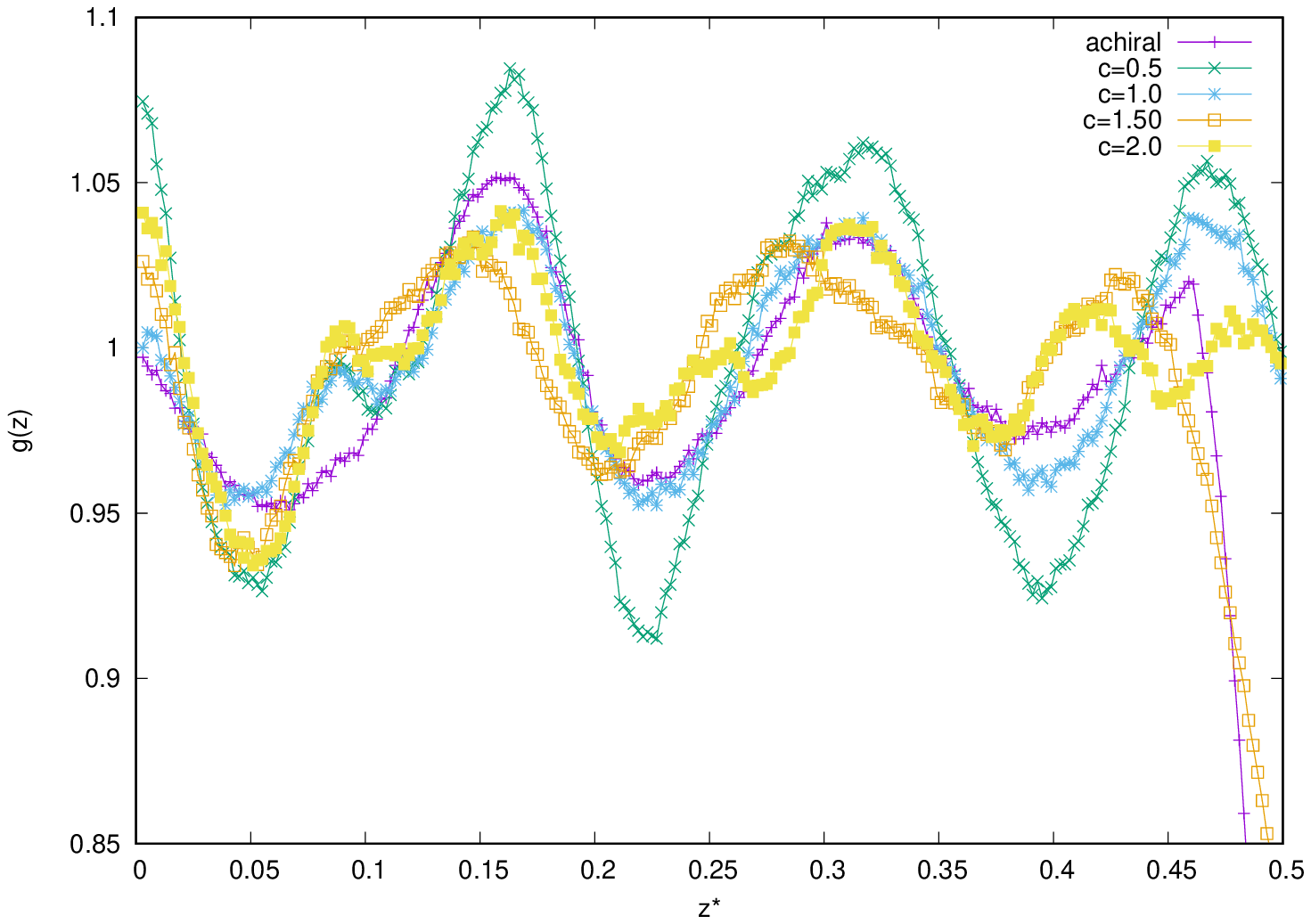}}
\caption{\label{fig:4} Plot of g(z) vs z* at different chiral strength}
\end{figure}

\begin{figure}
\subfloat[set-1 biaxiality]{\includegraphics[angle=-90,width=4.25cm]{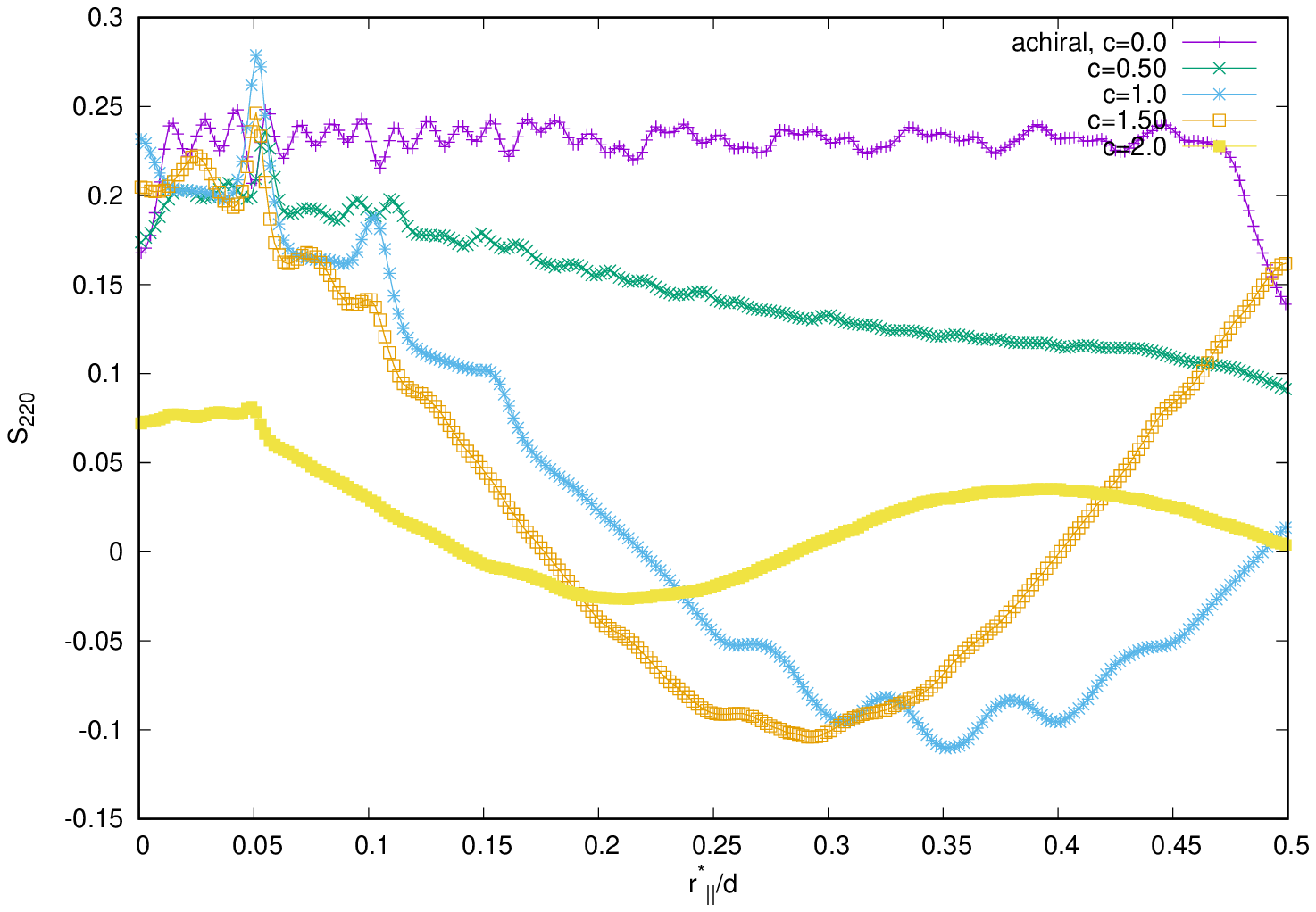}}
\subfloat[set-2 biaxiality]{\includegraphics[angle=-90,width=4.25cm]{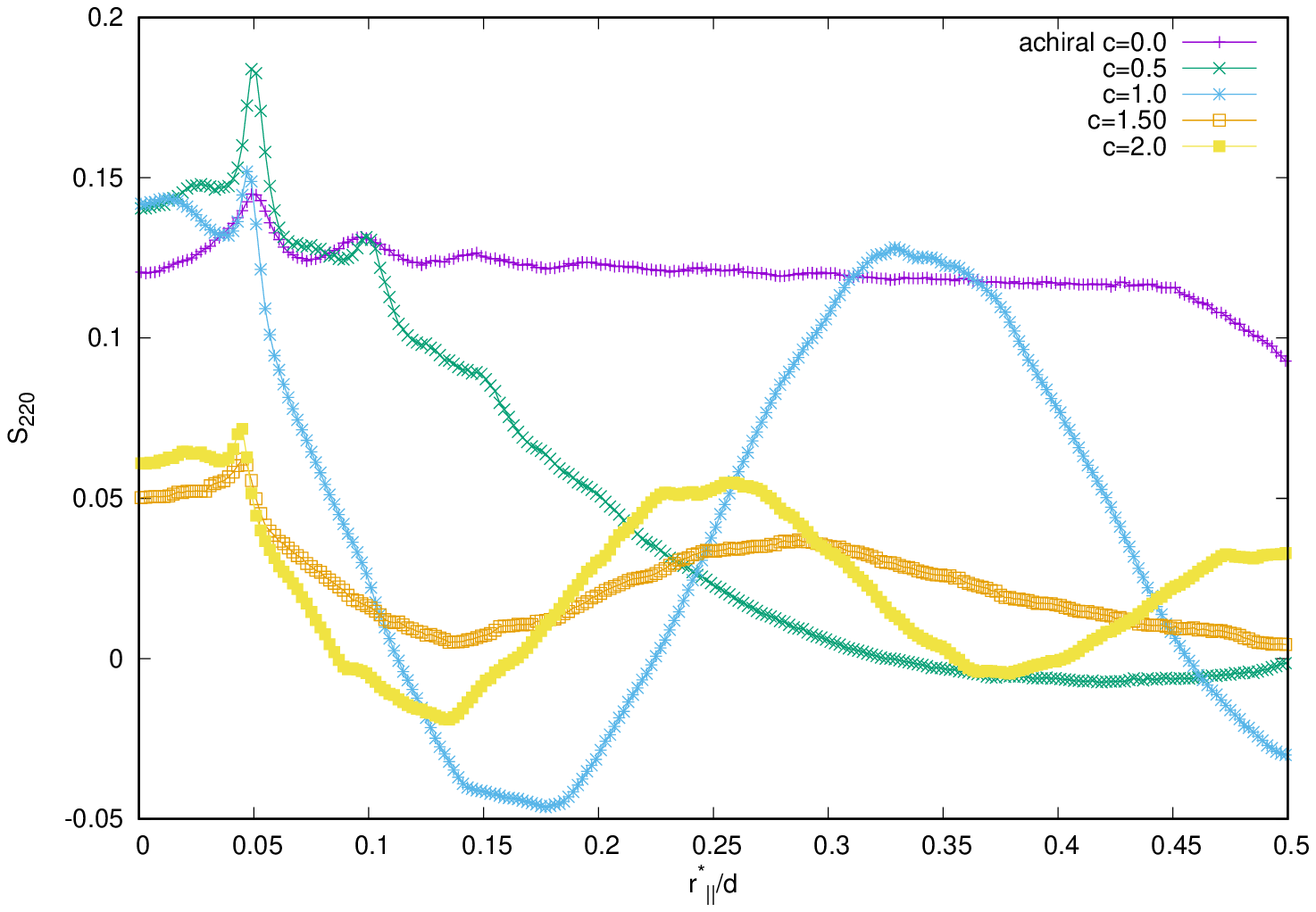}}
\caption{\label{fig:5} Plot of $S_{220}$ vs $r^{*}_{||}/d$ at different chiral strength}
\end{figure}

\begin{figure}
\subfloat[set-1 biaxiality]{\includegraphics[angle=-90,width=4.25cm]{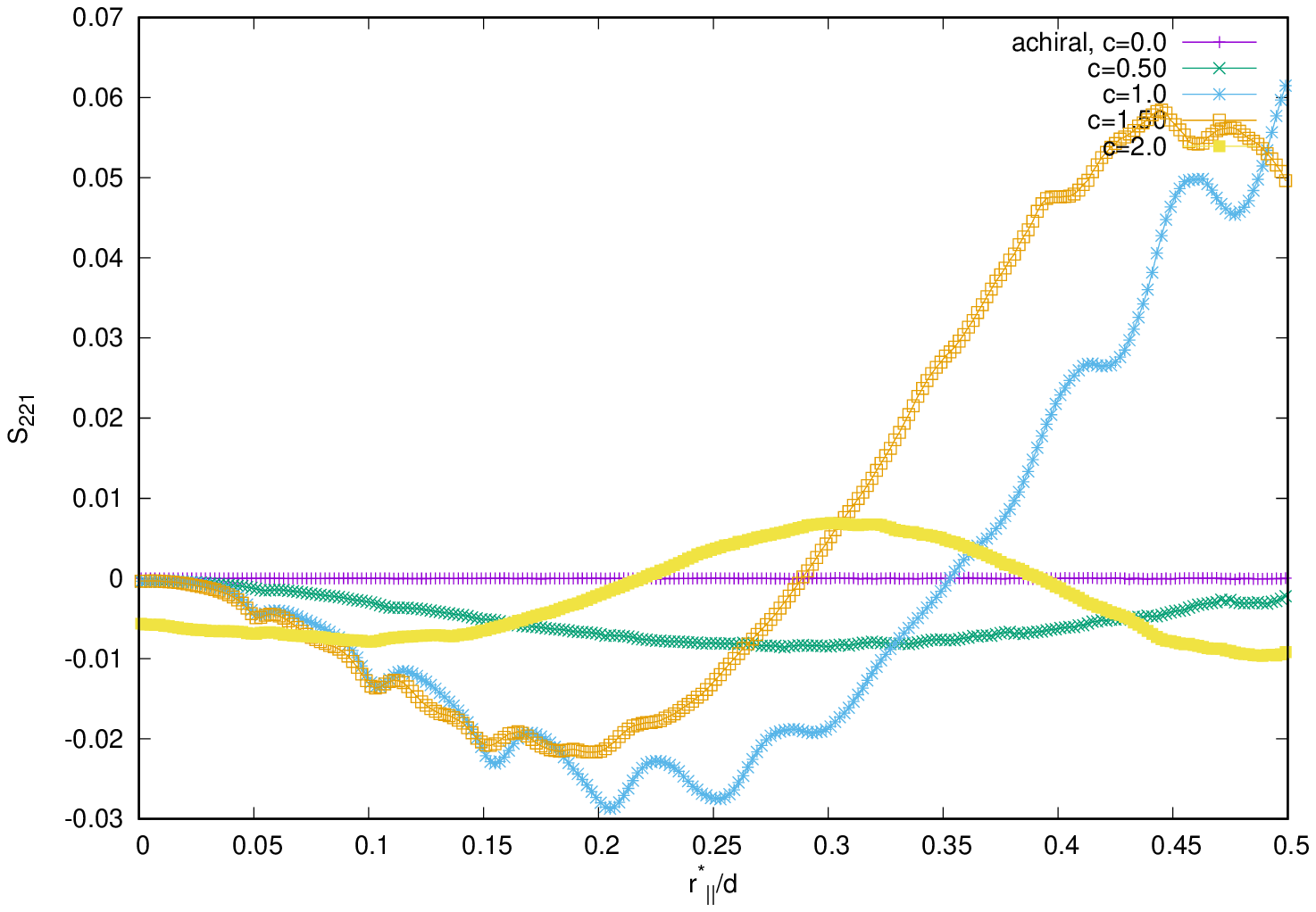}}
\subfloat[set-2 biaxiality]{\includegraphics[angle=-90,width=4.25cm]{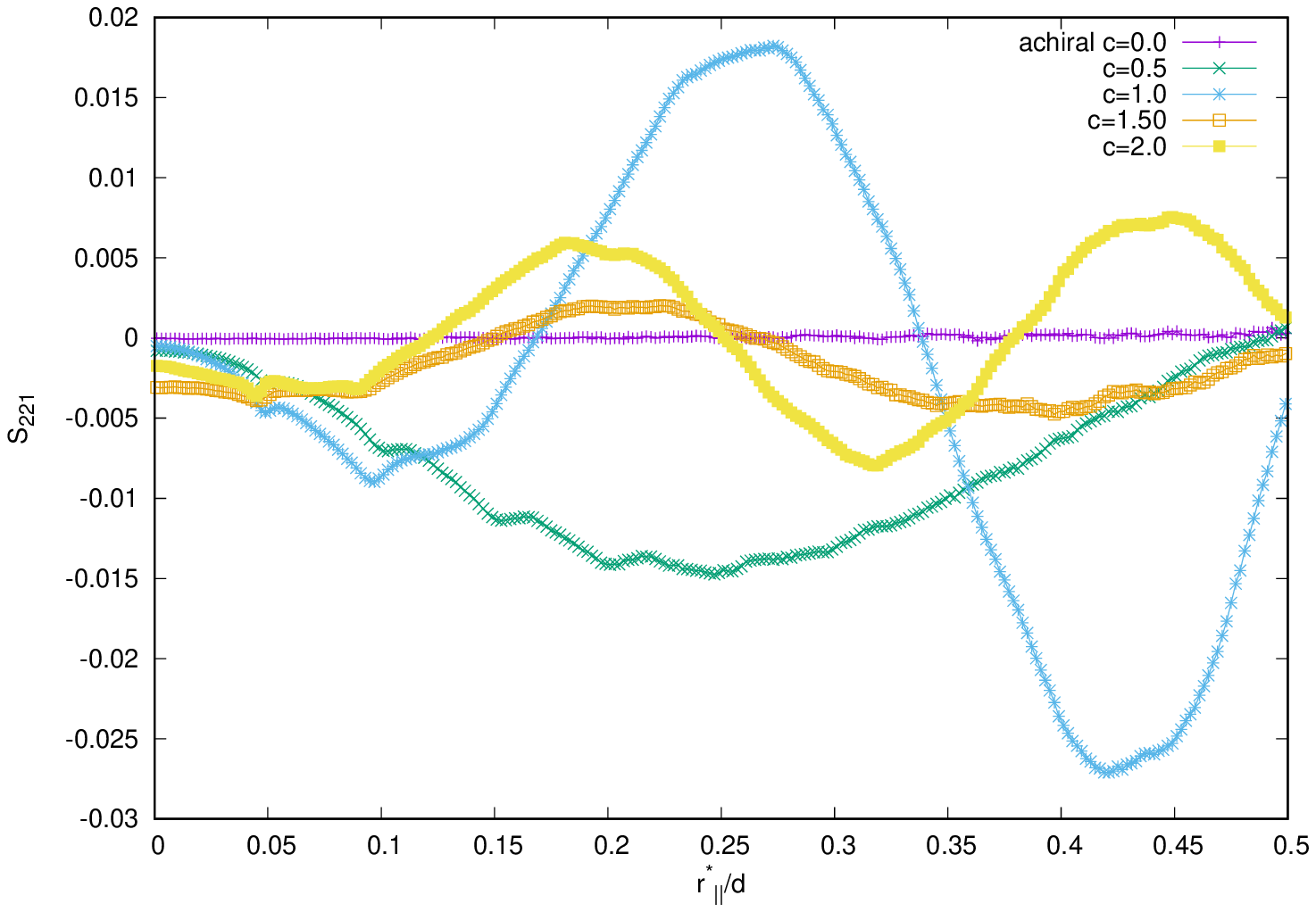}}
\caption{\label{fig:6} Plot of $S_{221}$ vs $r^{*}_{||}/d$ at different chiral strength}
\end{figure}

Molecular dynamics simulation of a system of size $N=1372$ molecular ellipsoids was done in the canonical ensemble(NVT) imposed with periodic boundary conditions. The values of well-depth parameters were taken to be $\mu=1$, $\nu=2$. We studied our system for two different sets of shape and energy biaxialities. For the biaxial molecules, the ratios $ \sigma_x \colon \sigma_y \colon \sigma_z $ were taken as $ 1 \colon 1.5 \colon 4.5 $ and $ 1 \colon 2.0 \colon 4.5 $ respectively for two different sets of shape biaxialities{\cite{Perram_et_al}}. Correspondingly values of $I_{xx},I_{yy},I_{zz}$ were taken as $(1.125,1.062,0.16)$ and $(1.212,1.062,0.25)$. Energy ratios of the biaxial molecules were $ \epsilon_x \colon \epsilon_y \colon \epsilon_z = 1 \colon (1.0/7.0) \colon (1/30.0) $ and $ 1 \colon (1/8.5) \colon (1/30.0) $ respectively for two different sets of energy biaxialities{\cite{saha3}}. In our system, initially the molecules had rotational and positional isotropy and they were located on an fcc lattice with minimum image convention inside the cubic simulation box. At the initial MD step, the translational and angular velocities were taken from the Gaussian distribution. To find the proper density{\cite{pasterny}} for closely packed phase structure for a given system size, we started our simulation run from a lower value of density $\rho^{*}=0.01$ with an increment of $0.001$ at a very low reduced temperature $T^{*}(\equiv {K_{B}T}/{\epsilon_0})=0.5$, till at the density $\rho^{*}=0.17$ and $0.13$ for two different sets of biaxialities for the given system size. We then melted the structure at $T^{*}=5.0$ to get the isotropic configuration. We used this equilibrated rotational and positional isotropic molecular arrangement as the initial configuration of our simulation studies in order to observe self-assembled phase behavior. For a particular temperature, the run started from a previous higher temperature equilibrated configuration, then we gradually decreased the reduced temperature $T^{*}$ to explore the phase structures formed by the system. The cut-off radius was taken as $R_{cut}=5.0\sigma_{0}$ and time-step size was $dt=0.001$ during the cooling process. We used leapfrog-verlet algorithm together with Nose-Hoover thermostat{\cite{holian}}. For the orientations of the biaxial ellipsoids we used quaternions, which is a set of four numbers describing the molecular orientations in 3-d Euclidean space. We simulated this system for different values of the chiral strength parameter $c$.
\begin{figure}
\subfloat[$c=0.50$]{\includegraphics[width=2.25cm]{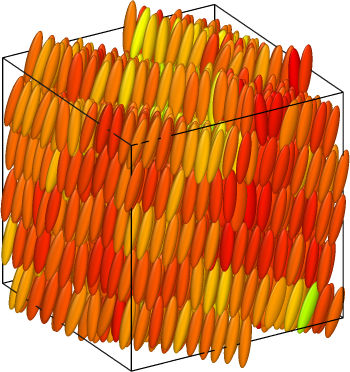}}
\subfloat[$c=1.0$]{\includegraphics[width=2.25cm]{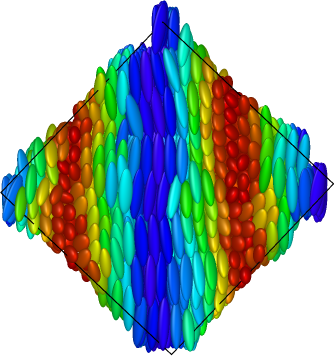}}
\subfloat[$c=1.50$]{\includegraphics[width=2.25cm]{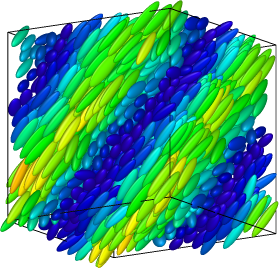}}
\subfloat[$c=2.0$]{\includegraphics[width=2.25cm]{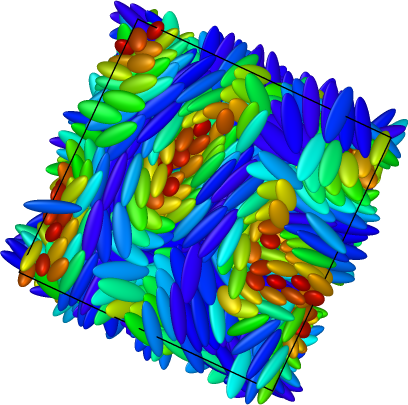}}
\caption{\label{fig:1} Snapshots of the configuration file at different chiral strength for $ \sigma_x \colon \sigma_y \colon \sigma_z = 1 \colon 1.5 \colon 4.5$ and $ \epsilon_x \colon \epsilon_y \colon \epsilon_z = 1 \colon (1.0/7.0) \colon (1.0/30.0) $ from ovito software.}
\end{figure}
\begin{figure}
\subfloat[$c=0.50$]{\includegraphics[width=2.25cm]{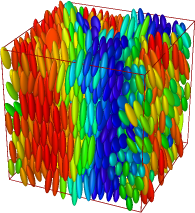}}
\subfloat[$c=1.0$]{\includegraphics[width=2.25cm]{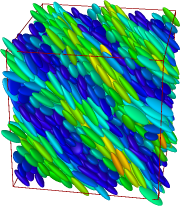}}
\subfloat[$c=1.50$]{\includegraphics[width=2.25cm]{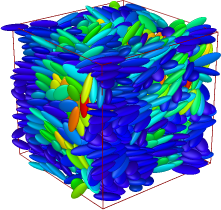}}
\subfloat[$c=2.0$]{\includegraphics[width=2.25cm]{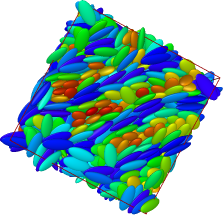}}
\caption{\label{fig:2} Snapshots of the configuration file at different chiral strength for $ \sigma_x \colon \sigma_y \colon \sigma_z = 1 \colon 2.0 \colon 4.5$ and $ \epsilon_x \colon \epsilon_y \colon \epsilon_z = 1 \colon (1/8.50) \colon (1/30.0) $ from ovito software.}
\end{figure}
\\The plot of pair correlation function $g(r)$ from Fig.~\ref{fig:3} shows short range positional order while there is no long range positional order. Another plot of density of molecular center of masses along the local director axis have been calculated in Fig.~\ref{fig:4}, which shows the formation of layers along the plane perpendicular to the local director. To investigate the orientational correlation between the molecules a scalar orientational longitudinal correlation function{\cite{Stone}} $S_{220}({r^{*}_{||}}/{d})$ and a pseudo-scalar orientational longitudinal correlation function $S_{221}({r^{*}_{||}}/{d})$ were calculated, which is a function of intermolecular distance ${r^{*}_{||}}$ measured along a reference axis which was further scaled by a distance $d$ related to the periodicity of the phase. The mathematical forms of the orientational functions are given by,
\begin{equation}
S_{220}({r^{*}_{||}}/{d}) = {\frac{1}{2\sqrt{5}}}{\langle 3(\hat{a}_{i3}\cdot\hat{a}_{j3})^{2}-1 \rangle}
\end{equation}
\begin{equation}
S_{221}({r^{*}_{||}}/{d}) = -\sqrt{\frac{3}{10}}{\langle \{(\hat{a}_{i3}\times\hat{a}_{j3})\cdot{\hat{r}_{ij}}\}(\hat{a}_{i3}\cdot\hat{a}_{j3}) \rangle}
\end{equation}
To calculate these functions, the averages were taken over $500$ different molecular configurations after reaching equilibrium at each $100$ steps, where a pair of biaxial molecular ellipsoids was separated by a distance $\frac{r^{*}_{||}}{d}$ along the chosen reference axis considering minimum image convention and scaling length $d$ which was equal to our simulation box length. The function $S_{220}$ was maximum when the semi-major axes of two molecules were parallel to one another, whereas $S_{221}$ obtained an extremum value when the semi-major axes of two molecules were at $45^{\circ}$ angle in between them. In order to investigate the influence of biaxiality on a system of chiral molecules, we calculated these correlation functions for two systems of different biaxialities and also for different values of chiral strength parameter $c$. The plots from the Fig.~\ref{fig:5} and Fig.~\ref{fig:6} shows flat correlation when the chiral strength parameter is $c=0$ which is the case of well known nematic phase and at $c=0.5$, there are very weak correlation inside our system. With the increment of chirality for both sets of biaxialities the values of both $S_{220}$ and $S_{221}$ show stronger correlations and the structures become more closely packed lowering the pitch value which can be seen from the graphs as more peaks are trying to fit within the box length. The orientational correlation has been measured along different box axes as well as director axes. The plots of Fig.~\ref{fig:5} \& Fig.~\ref{fig:6} have been calculated along x-director axis, while the correlations among other director axes as well as box axes shows no significant correlations which suggests that the biaxial ellipsoidal molecules are parallel to one another along a reference axis which acts as the helix axis forming cholesteric phase. 
For the orientational order parameters, we have calculated $\langle Q^{2}_{\alpha \beta} \rangle$ \& $\langle R^{2}_{2,2} \rangle$. The tensor order parameter for uniaxial phase is defined as, $Q_{\alpha \beta} = \frac{1}{2N}\sum(3{\hat{a}_{i\alpha}}{\hat{a}_{j\beta}} - \delta_{\alpha\beta}) ; \alpha,\beta = x, y, z$. Where, $\hat{a}_{i\alpha}$ is the $\alpha$'th component of unit vector $\hat{a}_{i}$ along the symmetry axis of the $i$th molecule. The average direction of the molecular allignment can be found by calculating the molecular director which corresponds to the eigenvector of the second rank tensor. 

\begin{figure}
\subfloat[set-1 biaxiality]{\includegraphics[angle=-90,width=4.25cm]{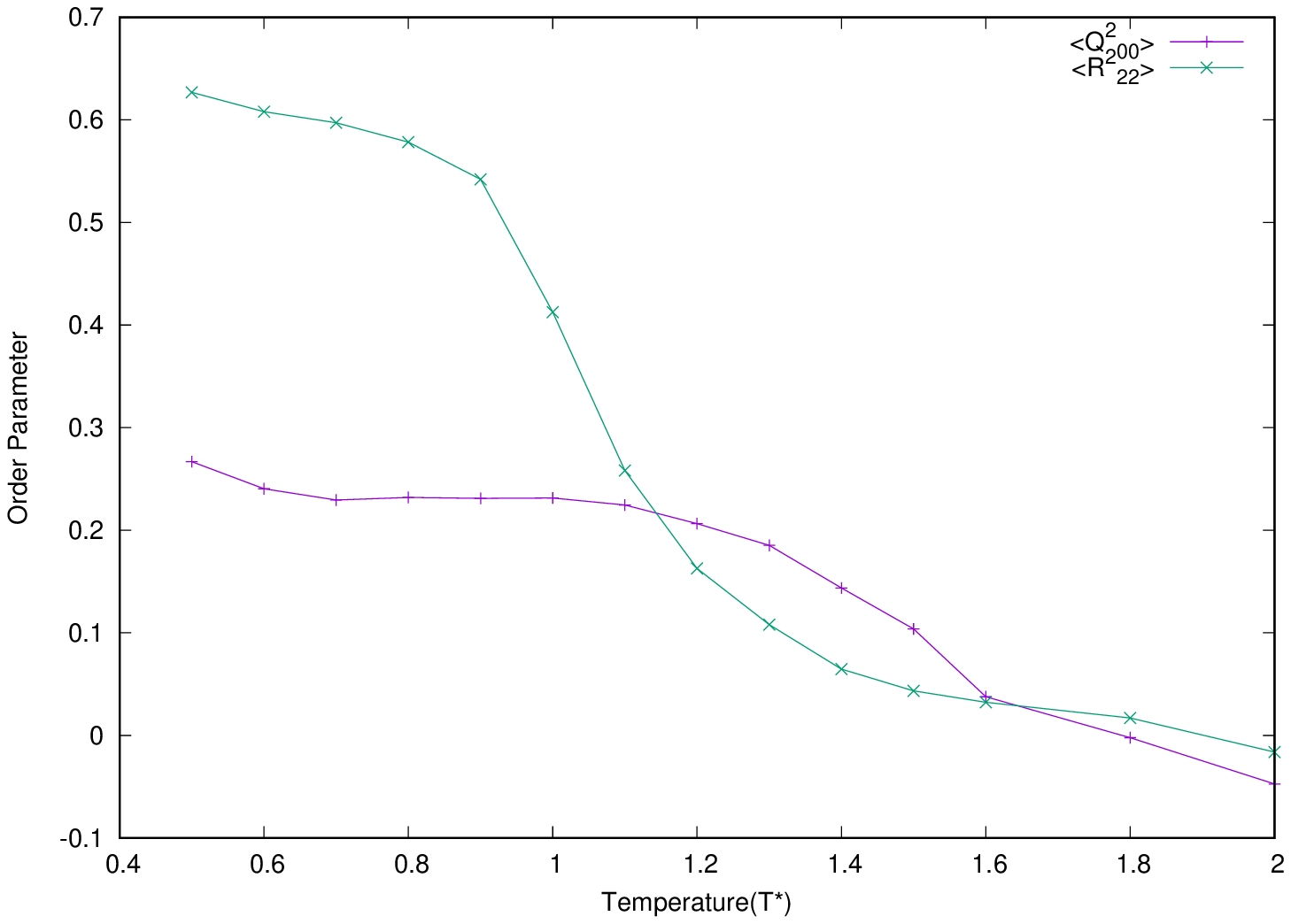}}
\subfloat[set-2 biaxiality]{\includegraphics[angle=-90,width=4.25cm]{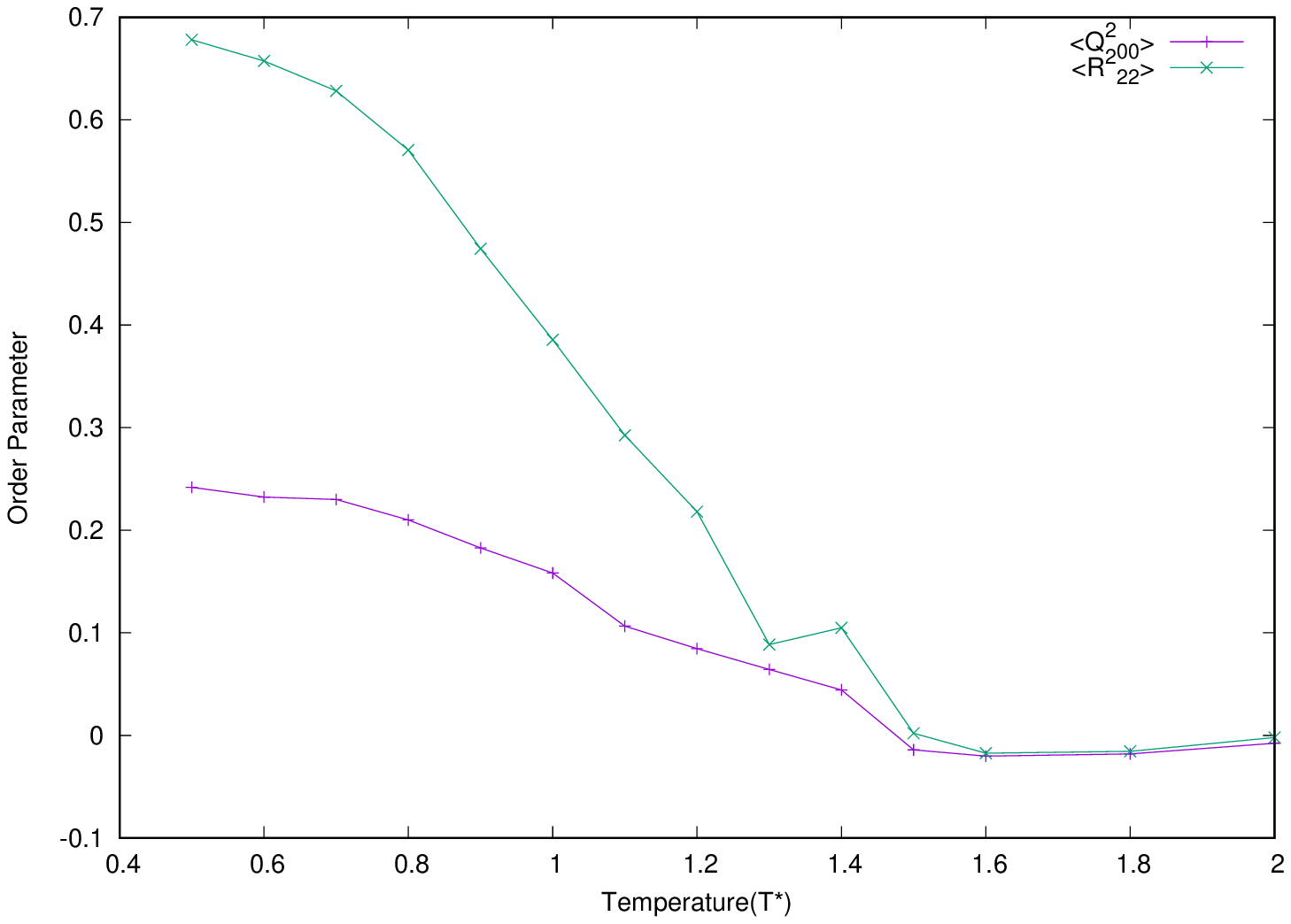}}
\caption{\label{fig:7} Plot of order parameters with temperature variation}
\end{figure}

The biaxial order parameter{\cite{Biscarini}} is $\langle R^{2}_{2,2} \rangle = \langle \frac{1}{2}(1+cos^{2}{\theta})cos2\phi cos2\psi - cos{\theta} sin2\phi sin2\psi \rangle$. Here, $\theta$, $\phi$, $\psi$ are the Euler angles calculated from the values of quaternions. As the temperature decreases the value of $\langle R^{2}_{2,2} \rangle$ increases which suggests that the phase biaxiality increases. This shows that higher the molecular biaxiality higher the phase biaxiality particularly evident at lower temperatures, is shown in the temperature variation plot of order parameters in Fig.~\ref{fig:7}. Mean-field studies{\cite{Alben,Straley}} of biaxial molecules showed that they tried to align their flat faces along a preferred orientation perpendicular to the director axis. From the snapshots of our system in Fig.~\ref{fig:1} and Fig.~\ref{fig:2} given above it can be verified. Higher molecular biaxiality means their flat face orientational order is also higher.

\section{Conclusion}

Biaxiality and chirality are two major aspects in technologically important material designing and formation of self-assembly in biological systems. Molecular biaxiality have a possible application in fast switching{\cite{Meyer_et_al}} electro-optic devices. One of the major component of cell membrane is cholesterol molecules{\cite{Cholesterol}}, which are biaxial in nature and also shows single handedness{\cite{Blackmond}}. In this simulation study, the effect of molecular biaxiality has been considered for a system together with chiral interaction extensively. When molecules deviate from uniaxial cylindrical symmetry then the shape as well as energy anisotropies are to be considered in lower symmetry cases. Phase biaxiality{\cite{Berardi}} is introduced by the molecular symmetry which is lower than the cylindrical symmetry and as a consequence, biaxial stacking can be seen at lower temperatures. In our computer simulation study the formation of biaxial cholesteric phase has been realized successfully. Due to lower molecular symmetry at higher chirality flat cylindrical domains are formed. When molecular biaxiality is higher the chiral pitch is decreased, which is consistent with the result of Yang et al{\cite{Yang2018}}. We hope our study may help in understanding the structure-property relationships of biaxial chiral liquid crystalline phases. Biaxial cholesteric liquid crystals have the properties of cholesteric helical structures together with biaxial ordering, which leads to its applicability in tunable optical properties and applications in photonics.

\section{DATA AVAILABILITY STATEMENT}
The data that support the findings of this study are available from the corresponding author upon reasonable request.

\begin{acknowledgments}
Sayantan Mondal wish to acknowledge the support of the Council of Scientific \& Industrial Research(CSIR), India for the Senior Research Fellowship(SRF) grant.
\end{acknowledgments}

\bibliography{aipsamp}

\end{document}